\documentclass[12pt]{iopart}

\usepackage[pdftex]{graphicx}
\usepackage{iopams}

\begin{document}

\title[Neutrino Experiments and LHC]{Neutrino Experiments and the LHC:\\
  Friends Across 14 Orders of Magnitude}

\author{J.M. Conrad}

\address{Physics Dept., Massachusetts Institute of Technology,\\ 77 Massachusetts Avenue, Cambridge, MA
  02139 US}
\ead{conrad@mit.edu}
\begin{abstract}
This proceeding explores some of the questions that connect the LHC and neutrino
experiments:  What is the origin of mass?  What is the
meaning of flavor?  Is there direct evidence of new forces or
particles?  The neutrino program investigating these questions is large
and diverse.  The strategy here, to narrow the discussion, is to focus
on relatively new ideas for experiments that may be less known within
the LHC community.

\end{abstract}

\maketitle

\section{Introduction}

Despite the wide difference in energy scales, the LHC and neutrino
experiments have a great deal of intellectual overlap.    This
talk explored three high-level questions as examples:
\begin{itemize}
\item What is the origin of mass?
\item What is the meaning of flavor?
\item Is there direct evidence of new forces or particles?
\end{itemize}
These are questions that resonate with the LHC community.
In this proceeding, I explore information and ideas that the neutrino
community adds to the debate.    

This discussion of the neutrino program has two biases used to narrow the scope to a
manageable scale for a 20 minute talk and 15 page proceeding.    First, the approach is data-driven.
A separate talk at the symposium, by Stephen Parke, was given on neutrino
theory, and the reader is referred to this for a more top-down
approach to the questions.   Second, the emphasis is on highlighting recent
experimental ideas which may be new to the LHC
community.   This necessarily leaves out a large number of exciting,
but better-known experiments,  however some very good reviews of these
are available in Refs. \cite{rev1, rev2, rev3, rev4, rev5}.

\section{The $\nu$SM}

The discussions below assume a ``neutrino Standard Model'' ($\nu$SM).
This is a minimal increment to the Standard Model
driven by the present $>5\sigma$ results.   This phenomenology has
been developed with an agnostic approach to the 
underlying theory.   It simply describes the data.

At this point, as demonstrated by LEP, we know that there are only three active flavors,  $\nu_e$,
$\nu_\mu$ and $\nu_\tau$, with masses less than $M_Z/2$ \cite{LEP}.
We know that these are related to at least three mass states, $\nu_1$,
$\nu_2$ and $\nu_3$, although there is not a one-to-one
correspondence.  In fact, the data are consistent with very large
mixings with the mass states \cite{mixresults}: 

\begin{equation}
\hspace{-2.5cm}
U_{PMNS} =
\left( \begin{array}{ccc}
 U_{e 1 } & U_{e 2}  & U_{e 3}   \cr
 U_{\mu 1  } & U_{\mu 2}  & U_{ \mu 3}   \cr
 U_{\tau 1  } & U_{\tau 2}  & U_{ \tau 3}   
\end{array} \right)
=
\left( \begin{array}{ccc}
0.795-0.846 & 0.513-0.585 & 0.126-0.178  \cr
0.205-0.543 & 0.416-0.730  & 0.579-0.808 \cr
0.215-0.548 & 0.409-0.725 & 0.567-0.800 
\end{array} \right), \label{pmns} 
\end{equation}

\noindent Reaching this level of accuracy, with 
every element of this Pontocorvo--Maki--Nakagawa--Sakata (PMNS) matrix
measured at some level, represents a highlight of the work
of the last decade in neutrino physics.

The limits on neutrino mass from kinematic studies of  tritium beta decay 
indicate that the neutrino mass states are less than $\sim 1$ eV \cite{beta}.
We know that at least two of the three mass states
must have non-zero mass, because we have measured 
two distinct mass splittings in oscillation experiments to high
accuracy in
atmospheric \cite{atmospheric}
and solar neutrino experiments \cite{solar}.   
Three neutrino mass states can be mapped onto two 
distinct splittings, and, when combined with reactor \cite{rev1} and
accelerator neutrino experiments \cite{rev2}, yield
$\Delta m^2_{31} = (2.473\pm0.069)\times 10^{-3}$ eV$^2$ and
$\Delta m^2_{21} = (7.50\pm 0.19) \times 10^{-5}$ eV$^2$
\cite{mixresults}.

These additions to the Standard Model (SM)-- that neutrinos mix as per
$U_{PMNS}$, and that at least two mass states are nonzero with mass
less than 1 eV--are assumed throughout the
discussion below.

\section{What Can Neutrinos Say About the Origin of Mass?}

With the discovery of the Higgs, we have made a major step forward in
understanding how mass terms should appear in the SM Lagrangian.
However, as highlighted at the conference, the underlying meaning of
the  fermion mass
spectrum we observe is unclear.   Moreover,
we have yet to find any indication for a  mechanism which 
prevents the masses of all fermions from being at the Planck 
scale.    So, obviously, something is
seriously wrong.  We need more clues.

A place to look for more clues is the neutrino sector.
Other than the facts that at least two neutrino states must have non-zero
mass and that the mass spectrum corresponding to the active flavors must be less than $\sim$1 eV,  we know very little about
neutrino masses.      Our limited knowledge raises
a host of other questions:   How far below the upper limits do the
neutrino masses lie?    Why would the
coupling of the Higgs to the neutrinos be more than five orders of
magnitude less than the couplings to the charged fermions?     Can
there be other mass-producing mechanisms at play that lead to this
effect?   And will neutrinos have the same mass hierarchy as the quark
sector,  with the
small splitting seen at the lowest masses and the large splitting
associated with the highest mass?    These are all questions we can
investigate in the next decade to shine more light on the continuing
question of the origin of mass.

At present, we know that the gap between the neutrino mass states of
the $\nu$SM and the
electron mass is five orders of magnitude, This is as large as the
span of masses of the charged fermions, from electron to top quark.  
However, this is just a limit in the neutrino sector,
and the lightest mass state might be as low as  $\sqrt(\Delta
m^2_{31}) \sim 50$ meV, leading to a ``desert'' between the fermions 
of $10^7$ eV.   As we think about the problem of mass, we must
consider what produces such a gap.    There are two opposing
approaches.  One introduces new physics into the $\nu$SM, such
as the See-saw Model \cite{seesaw} to motivate the gap.   The other
argues that the masses are just an accident of nature, like the orbits
of the planets, and, as in the case of orbits of planets, gaps happen.
However, to push the analogy further,   the study of the Mars-Jupiter gap
has given interesting insights into the formation of the solar system
and potential exo-solar-systems;  similarly,  even if the specific values
of our fermion mass spectrum turn out to be accidents of nature,  the origin of 
this ``gap-feature'' may lead to interesting insights in particle physics.

The first question, then, is:  How big is the gap?  The most
precise method of attack comes from the study of the kinematics of 
tritium $\beta$ decay \cite{beta}.   Neutrino mass will lead to a
lower endpoint of the $\beta$-decay spectrum.  These
experiments measure $m_{\nu_e}$, which is a flavor weighted
average of the neutrino masses: 
\begin{equation}
m_{\nu_e}^2 = \sum |U_{ei}|^2 m_{\nu_i}^2, \label{mnue2}
\end{equation}
where the sum, $i=1..3$, is over the three mass states.   
The allowed values of $m_{\nu_e}$ in nature 
depend upon 1) the absolute offset of the neutrino mass states,
which is the equivalent of the mass of the lowest state, and 2)  the mass hierarchy.

\begin{figure}[tb]
\includegraphics[scale=0.5]{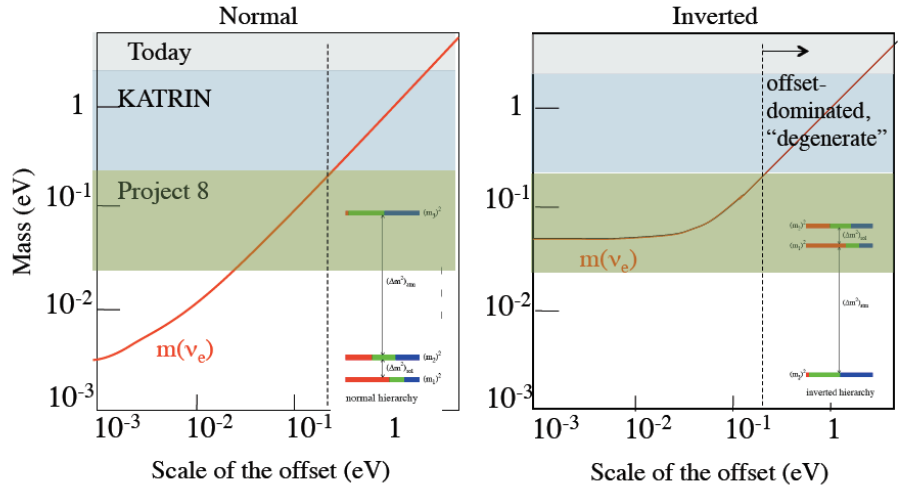}
\vspace{-3.cm}
\caption{Cartoon of the possible values of $m_{\nu_e}$ as a function of 
absolute offset. Left and right apply to the two hierarchies,
shown in inset.   The KATRIN and Project 8 sensitivities are shown.
\label{direct}}
\end{figure}

To illustrate the point, consider Fig.~\ref{direct}, which is a
cartoon of possible values of $m_{\nu_e}$ as a function of 
absolute offset, where left and right apply to the two hierarchies,
shown in inset.  If the absolute offset is large with respect to the 
mass splittings, then there is little difference between the
potential  $m_{\nu_e}$ values for the two hierarchies--this is called
the ``degenerate range.''    However, for smaller values of offset,
there is a substantial difference in the range of possible values
which must be probed.   To see this,  consider the inset diagrams, 
where each bar represents a mass 
state and the colors indicate the flavor mixings.    The mass state 
$\nu_3$ is defined to have the smallest electron
flavor content.   The normal hierarchy (left) places $\nu_3$ at the
top of the mass spectrum, thus the highest mass state contributes a
small weight in Eq.~\ref{mnue2}.  On the other hand,  the inverted hierarchy places $\nu_3$ at
the bottom,  resulting in large electron-flavor content in two high
mass states.        This inverted  arrangement allows an experiment with
sensitivity below $m_{\nu_e} \sim 0.05$ eV to cover the entire range
of potential values.

KATRIN, which will run in 2015, will be the first experiment to weigh
in, and will have a sensitivity of $\sim 0.2$ eV at  90\% CL\cite{KATRIN}.  This will cover the
degenerate range of potential solutions.     KATRIN is a classic
electromagnetic spectrometer.  To reach high resolution at the
$\beta$-decay endpoint,  the central
region of the spectrometer must be 10~m in diameter, which is
enormous and is likely to make the KATRIN experiment the last of its
kind.

To move to the next order of magnitude in sensitivity a new technology
is required, and an interesting
possibility has been put forward by the Project 8 collaboration \cite{P8}.
This technique traps the $\beta's$ from tritium decay in a magnetic 
bottle.  As the electrons traverse the bottle,  they will radiate in
the RF, at the fW level.  In principle, the radiation can be observed
with MHz antennae now under development as listening devices for cell
phones.  The combination of time-of-flight and the frequency of the
radiation allows the electrons with energies at the very endpoint of
the decay to be isolated and counted.     This has the potential to
push the sensitivity to $m_{\nu_e}$ down to $\sim 0.02$ eV, covering
  the entire range of potential values in the case of the inverted hierarchy.

A related question to the absolute mass offset is whether neutrinos
acquire mass in the same way as the other fermions.   Because
neutrinos are neutral with respect to the electromagnetic and strong
forces, they can, potentially, be their own antiparticle--where
neutrino and antineutrino are distinguished by the spin state.  
This allows introduction of an additional ``Majorana'' mass term, beyond the Higgs
mechanism, into the Lagrangian.   Through the See-saw model, this mass
term can be connected to physics at higher energy scales, leading to
an explanation of the large mass gap we observe.

The most precise way to test for the Majorana nature of neutrinos is
through neutrinoless double beta decay ($0\nu\beta\beta$).    This is
the neutrinoless analogue to the observed process of double beta decay to 
$\beta \beta \bar \nu_e \bar \nu_e$, which has been observed to occur
in the handful of elements where single beta decay is energetically
forbidden.   In the case of $0\nu\beta\beta$,  the two antineutrinos
annihilate--allowed by their Majorana nature.   While it might be surprising
to think that total lepton number can be violated in this way,  in
fact nothing in the SM prevents this.   So under the argument that,
``if is it not forbidden, it is compelled,''   
$0\nu\beta\beta$ is natural.

\begin{figure}[tb]
\begin{center}
\includegraphics[scale=.4]{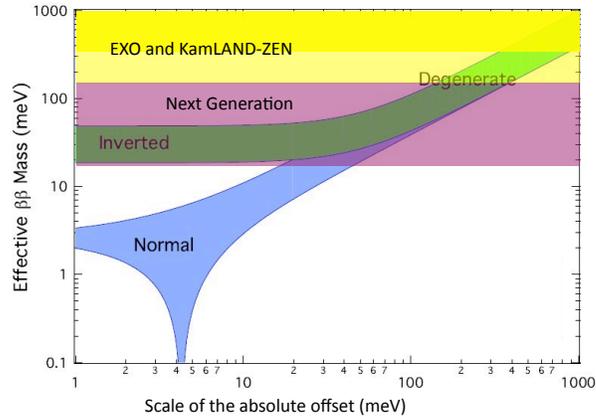}
\end{center}
\vspace{-2.cm}
\caption{The allowed values of $m_{\beta \beta}$ as a function of absolute mass
offset for the inverted and normal
mass hierarchies, overlaid.  The uncertainty from the matrix element is
indicated by the pale yellow region.   Next generation experiments are
described in the text.
\label{mbbplot}}
\end{figure}

The signal for $0\nu\beta\beta$ is the production of two monoenergetic
electrons at the end-point of the $\beta$ spectrum.   Thus these
experiments have a great advantage in knowing exactly where to look
for a new physics signal.   The expected rate is related to a
flavor-weighted neutrino mass:
\begin{equation}
|\langle m_{\beta\beta}\rangle| = | \sum m_i U_{ei}^2|. \label{mbetabeta}
\end{equation} 
The allowed values of $m_{\beta \beta}$ as a function of absolute mass
offset are shown in Fig.~\ref{mbbplot} for the inverted and normal
mass hierarchies, overlaid.    The basic form is similar to that of
$m_{\nu_e}$ from the direct mass searches, with a degenerate region
for large absolute offsets and larger average mass expected for
inverted rather than normal hierarchy.   But the differences in
flavor-weighting (compare Eqs.~\ref{mnue2} and \ref{mbetabeta}) lead
to a spread of potential allowed values in the $0\nu\beta\beta$ case arising
because the elements of $U$ can have arbitrary phases.   As a result,
the allowed regions are wide bands on Fig.~\ref{mbbplot}.   An accurate
measurement of $m_{\nu_e}$ can allow us to hone in on these
$CP$-violating phases if $0\nu\beta\beta$ is observed, providing
valuable input to Leptogenesis models \cite{Leptogen}.

The progress in the search for $0\nu\beta\beta$ is indicated by the
yellow shaded region, which are the limits from EXO and KamLAND-Zen 
\cite{rev3}.   The excluded region becomes pale in the lower 
regions, indicating the theory error from the calculation of the
nuclear matrix element of Xenon.       The theory error for all of the
potential $0\nu\beta\beta$ elements is large.  This has led to
a set of next generation $0\nu\beta\beta$ experiments that employ
a range of different elements, so that cross comparison of signals and
limits can allow a precise interpretation of the results with less
sensitivity  to the underlying nuclear theory.   The elements include 
Neodymium (SNO+), Tellurium (CUORE, potentially SNO+), Germanium (GERDA and
Majorana) and Xenon (NEXT, as well as continuations of EXO and 
KamLAND-Zen) \cite{rev3}.     

As with the direct mass measurements, the ambition of the next
generation is to entirely cover the potential values of
$m_{\beta\beta}$ for the inverted mass hierarchy.    Each of the above
experiments is pressing for improvements to reach this level, and it
is unclear, which, if any, will succeed.    However, an interesting
new step is being pursued by SNO+,  to switch from Nd to Te, 
which may make this experiment the
first to pass below the inverted hierarchy in sensitive.    This step
can be taken because of a very nice synergy among neutrino
experiments.   A set of recent reactor-based experiments has solved
the problem of doping scintillator oil with a high fraction (a few
percent) of metal isotopes \cite{Yeh}--research pursued to improve the neutron
capture cross section in those experiments.   This same technology
appears to allow SNO+ to dope with 3\% Te.   Since the natural
abundance of the double beta decaying isotope of Te is 34\%,  this
results in sensitivity across the full range of potential mass values 
for SNO+, assuming they can achieve
the necessary resolution at the endpoint \cite{Klein}.

From the above discussion, it is clear that the hierarchy is playing a
crucial role in accessing the physics.   Beyond this, the hierarchy
itself is an interesting question, if one is seeking to make a model
of fermion masses.   Thus,  it would be best if the question of the
hierarchy could be addressed separately from the quantitative mass
measurements.  Luckily, this can be done in certain neutrino oscillation
experiments.

To understand the sensitivity of oscillations to the hierarchy,
consider the three neutrino mass states propagating through the
earth.   Because the earth is filled with electrons, the neutrinos
feel a weak potential.  The effect of this potential on the propagation
will be different for each mass
state, because of the varying electron-flavor content, leading to a
change in the oscillation probability.  This ``matter effect'' is enhanced with
energy and distance.  The sign of the matter effect is opposite 
for neutrinos and antineutrinos, and thus can be
regarded as faux-$CP$-violation.    But unlike true $CP$
violation, the effect will appear in both appearance and disappearance
oscillation experiments.

From the above description, one can see that an ideal setting to search
for matter effects, and thus determine the mass hierarchy, is long
baseline oscillations.   While the traditional approach has been to
look to accelerator-based sources,  if what one wants is extremely
long baseline, with a high-rate of events in the 1 to 20 GeV range,
then nothing beats a cross-earth atmospheric neutrino experiment.
To this end,  the IceCube Experiment is upgrading their Deep Core
central region with additional strings of PMTs in order to explore
this physics.   This upgrade, called PINGU, can be completed on a
relatively short timescale, and will have 3 to 5$\sigma$ capability
within a few years of running \cite{pinguwhitepaper}.   As a result,
one can imagine the mass hierarchy question--normal or inverted--
being answered on the same timescale as the direct and
$0\nu\beta\beta$ mass measurements.

The combination of the three approaches to questions of mass is powerful.
In the cases where both the direct and $0\nu\beta\beta$ experiments
see signals, very valuable information can be added to the models for
new physics at high mass scales, including Leptogenesis.    With or
without a signal observed in direct and $0\nu\beta\beta$ experiments, 
the result can constrain cosmology.   It should be noted that
cosmology gets a good fit assuming no neutrinos \cite{paper16}, and
mechanisms have been put forward that reduce or eliminate the cosmic
neutrino background \cite{rev4}, and so constraints from earth-based experiments
are quite important.  Lastly,  there is the
potential for experimental discrepancies that force us to entirely
rethink our nascent understanding of neutrino mass.

\section{What Can We Learn From Neutrino Flavor Studies?}

If one is looking for patterns in the SM, then the neutrino flavor
mixings expressed by Eq.~\ref{pmns} are as strange as the neutrino masses.   
Completely opposite to the quark sector,  all of the off-diagonal entries in the
mixing matrix are large.   Even the smallest entry,  $|U_{e3}|$ is an
order of magnitude larger than its quark-sector equivalent.  As with
the case of the masses, it may be that this is just a random
occurrence in nature--this model is called ``Anarchy'' in the neutrino
community.     But we are, at this point, far from the level of
precision where it is time to just give up on searching for a
pattern.  Indeed,  neutrino physics is, now, at the level of precision
of the quark sector in 1995 \cite{ckm95}:
\begin{equation}
U_{CKM}^{1995} =
\left( \begin{array}{ccc}
0.9745~to~0.9757 & 0.219~to~0.224 & 0.002~to~0.005  \cr
0.218~to~0.224 & 0.9736~to~0.9750  & 0.036~to~0.046 \cr
0.004~to~0.014 & 0.034~to~0.046 & 0.9989~to~0.9993 
\end{array} \right)\nonumber
\end{equation}
So a whole world of precision flavor measurement is only now opening up to
the neutrino community.  

\begin{figure}[tb]
\begin{center}
\includegraphics[scale=.6]{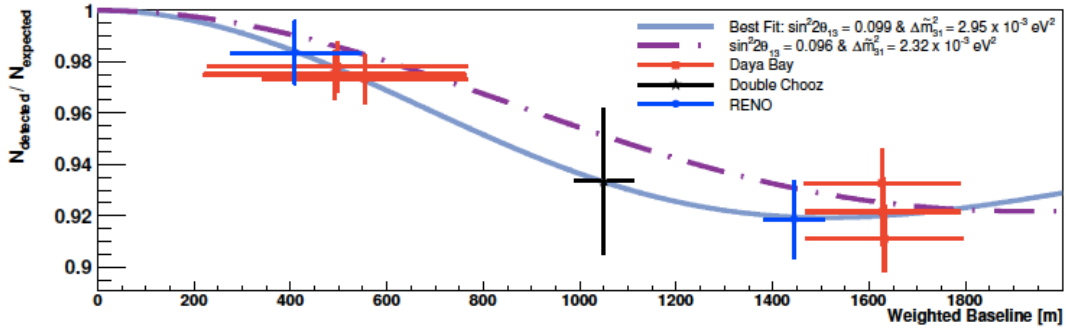}
\end{center}
\vspace{-6.5cm}
\caption{Measurements from the reactor experiments at various
  baselines, overlaid with a fit to $\Delta m^2$ \cite{Thiago}.
  Improved measurements of $\theta_{13}$ at multiple baselines will
  allow tests of deviations from the oscillation expectation due to
  non-standard interactions.
\label{LE}}
\end{figure}

The ultimate goal will be to develop unitarity tests that are as
precise as those in the quark sector.   However, this will require
precision measurement in muon-to-tau and electron-to-tau appearance
experiments.    Because of the tau mass suppression in charged current
interactions, this requires high energy neutrino beams.   Since
the oscillation length of these studies is fixed in the $\nu$SM by 
$\Delta m^2_{21}$ and $\Delta m^2_{31}$,   high energy inevitably
means ultra-long baselines, which require ultra-high intensity.   Thus
the only practical solution is a $>20$ GeV neutrino factory, which
will not be realized until far in the future.

Nevertheless, improving the precision measurements we can do today can
potentially produce indications of new physics.    The neutrino
community is very interested in models with non-standard interactions
that produce instantaneous transmutation of neutrino flavor at
production and interaction, as well as modify the oscillation
probability.   Cross comparing results of matrix element measurements
from different experiments may provide sensitivity to such effects.
But before looking for something completely different than in the
quark sector,  one can also look closely at the $\nu$SM to ask if it is complete.
We know that the quark mixing matrix has an arbitrary phase which
leads to $CP$-violation.  It is important to ask if such a phase is
appearing in the lepton sector also.   It would be quite striking if
it did not, since this would speak against the dictum of ``that which
is not forbidden is compelled.''   It would also be striking if the
value were large, the opposite to the quark sector, as that would make
the apparent dichotomy between the quarks and leptons even more
sharp.    Thus, even if unitarity tests are far away,  there are a
great deal of important checks we can pursue in neutrino flavor
physics in the near future.

As an example of a precision B$\nu$SM search that we can do within the 
next decade,  consider $\theta_{13}$.
Since the PMNS matrix simply produces a rotation between flavor and 
mass states,  the flavor mixing is most commonly parametrized as
through  three Euler angles, $\theta_{12}$, $\theta_{23}$ and
$\theta_{13}$.   The relationship between the matrix elements and the
angles is complicated except for one entry,  $U_{e3}$, which depends
purely on $\sin(\theta_{13})$.    This turns out to be the 
smallest mixing angle, and was
only recently observed to be non-zero \cite{mixresults}.  This was an exciting result both for 
theories describing the PMNS matrix and also because non-zero
$\theta_{13}$ is crucial for the potential to observe $CP$-violation.   
Neutrino physicists consider  2011-12 ``The Year of $\theta_{13}$'' for the neutrino
sector, in the same way it was ``The Year of the Higgs'' for the LHC.
And, now, like the Higgs,  the next thing to do is explore this new
measurement for hints of the unexpected.

What has been reported by the Double Chooz \cite{DC}, Daya Bay
\cite{DB} and RENO \cite{RENO}
reactor experiments, was 
a deficit in anti-electron neutrinos at $L/E \sim 1/\Delta
m^2_{tam}$.    This would be consistent with the expectation of
disappearance in the $\nu$SM model, and the mixing angle can be
extracted from this equation:
\begin{equation}
P(\bar \nu_e \rightarrow \bar \nu_e) \approx 1 - \sin^2 2\theta_{13} \sin^2
(1.27 \Delta m^2_{atm} L/E).
\end{equation}
The approximation arises from dropping subleading-terms from $\Delta
m^2_{12}$ which are very small and by
employing the assumption that $\Delta m^2_{13}=\Delta m^2_{23} =
\Delta m^2_{atm}$.

One sees immediately that the $\nu$SM 
makes very specific predictions as a function of $L/E$.  These are
modified in the presence of non-standard interactions.   And so an
important next step is to test for this $L/E$ dependence.
We can already begin to test the $L$ dependence of the $\theta_{13}$
measurements, because the three reactor experiments are at different
baselines \cite{Thiago}.     Fig.~\ref{LE} shows the data associated with the
various baselines.   Daya Bay has a particularly complicated
reactor-core to detector arrangement, and that is why they report
results from many baselines.   The dot-dashed line shows the
expectation from the $\nu$SM, allowing $\sin^2 2\theta_{13}$ to float
and fixing $\Delta m^2_{atm}$ to the measurement from MINOS \cite{MINOS}.  The
solid blue line allows both $\sin^2 2\theta_{13}$ and $\Delta
m^2_{atm}$ to float.  The result is a rather high value for $\Delta
m^2_{atm}$, but allowable within errors.    However, there is clearly
room for models beyond the $\nu$SM, introduced as sub-leading 
additions, to fit the data.   

Understanding $\theta_{13}$ is also key to the search for $CP$
violation in the neutrino sector.     $CP$ violation will modify the
oscillation probability of appearance experiments.    For
muon-to-electron appearance, which is a channel we can test in the
near future, the oscillation probability is given by:

\begin{eqnarray}
P  &  =&\sin^{2}\theta_{23}\sin^{2}2\theta_{13}\sin^{2}\Delta_{13}
\nonumber \\
&  & \mp\sin\delta_{cp}\sin2\theta_{13} \sin2\theta_{23}\sin2\theta_{12}\sin^{2}\Delta_{13}\sin\Delta_{12} \nonumber\\
&  &+\cos\delta_{cp}\sin2\theta_{13} \sin2\theta_{23} \sin2\theta\sin\Delta_{13}\cos\Delta_{13}\sin\Delta 
_{12}\nonumber\\
& & +\cos^{2}\theta_{23}\sin^{2}2\theta_{12}\sin^{2}\Delta_{12}%
,\label{equ:beam}%
\end{eqnarray}
where $\Delta_{ij}=1.27 \Delta m_{ij}^{2}L/E$, and $-(+)$ refers to
neutrinos (antineutrinos).

For the large values of 
$\theta_{13}$ which have been observed,  measurement of a 
non-zero $CP$-violating phase in the PMNS matrix
(eq.~\ref{pmns})  ensures non-vanishing baryon asymmetry \cite{Leptogen}.
The theoretical problem is how to quantify the effect so as to
understand whether this is the sole source of baryon asymmetry.
Thus, $CP$ violation is a place where experiment is pushing theory as hard as
theory often pushes experiment.     Two outcomes are particularly interesting:  if
experimentalists provide a measurement of $CP$ in the neutrino
sector that is large,  a major contribution to the 
baryon asymmetry is necessary\cite{Leptogen}; and if the $CP$ violation
parameter is limited to less than 5$^\circ$,  theorists must begin seriously
considering how to explain zero $CP$ violation in the lepton sector,
when it is observed in the quark sector.

\begin{figure}[tb]
\begin{center}
\includegraphics[scale=.4]{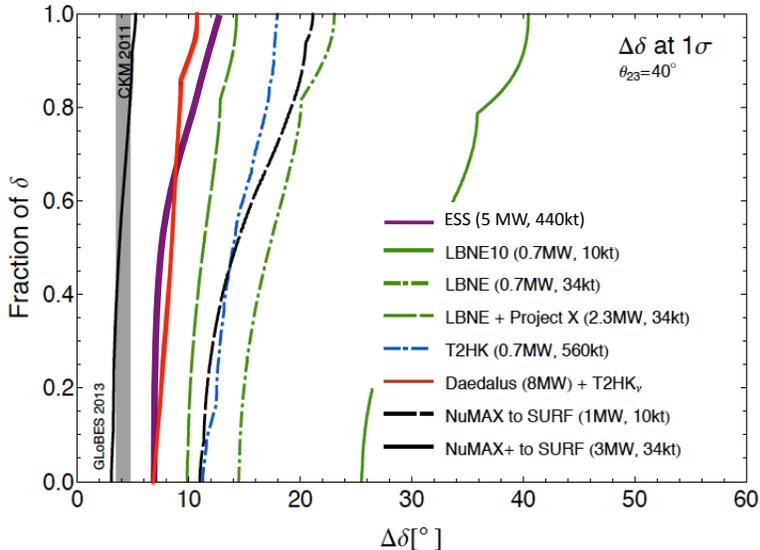}
\end{center}
\vspace{-1.5cm}
\caption{Fraction of $\delta$ parameter space covered at 1$\sigma$ 
  given the measurement precision of $\delta$.  The plot is from
  P. Huber's review at Snowmass 2013 \cite{Huber}, with the ESS
  expectation for the 360 km site added.   Conventional designs are in
  green, blue and purple.  DAE$\delta$ALUS, in red, uses three sites at
  1 MW, 2 MW and  5 MW to produce DAR beams to trace the oscillation
  wave.  The black line indicates the capability of a neutrino factory.
\label{ESS}}
\end{figure}

$CP$ violation necessarily requires an
appearance experiment, and the most feasible channel is muon to
electron flavor oscillations.
The classic, or ``conventional,'' approach to studying $CP$ violation is to exploit the
change of sign in the oscillation probability by running with
neutrinos versus antineutrinos (see Eq.~\ref{equ:beam}).   However, an
alternative method is to run strictly in either neutrino or antineutrino
mode, and instead trace out the oscillation wave, which is modified by
a nonzero value of $\delta$.    Because non-standard
interactions may also be occurring in the neutrino sector,
measurements using both approaches is warranted.
In either case,
a precise measurement of $CP$ violation is best done at low energies
and relatively short baselines, where matter effects do not add
ambiguity to the result.     

The newest and most powerful proposal for a conventional $CP$ violation experiment 
using
neutrino and antineutrino muon-to-electron flavor oscillations is to 
be built in Sweden.
This experiment \cite{ESS} would make use of the European
Spallation Source (ESS), now under construction at Lund,  
to produce a conventional decay-in-flight, wide-band neutrino beam 
peaking at about 200 to 300 MeV.  Operation of the ESS linac proton beam will start at reduced power in 2019, increasing to the full design power of 5 MW in 2022.
To produce the neutrino beam, the 2.5 GeV linear accelerator would be
upgraded by 
another 5~MW, allowing $\sim$10$^{23}$ protons per
year for neutrino production, 
concurrent with the neutron spallation running. 
A large water tank Cherenkov detector can be located underground in a
mine at a depth of $\sim$3000 mwe at two potential locations:
Zinkgruvan, which is 365~km from Lund, 
and Garpenberg, which is 540~km from Lund.   These locations offer very
similar rock and depths to the Pyh\"asalmi mine in Finland that is
under consideration for a CERN long baseline program \cite{mine}.  Data would be taken 
with 2 years of neutrino running and 8 years of antineutrino running.
The design of the water Cherenkov detector is that of MEMPHYS \cite{MEMPHYS}, which is 440 kt of water.

The resulting capability compared to other conventional beam
experiments is shown in Fig.~\ref{ESS}, which shows the fraction of 
$\delta$-space covered at 1$\sigma$ for a given precision in $\delta$.  
The ESS experiment (purple) substantially outperforms the 
other 
proposed conventional designs (LBNE (green)\cite{LBNE} and T2HK
(blue) \cite{T2HK}) due to several factors.  First, the low energy 
of the ESS beam highly suppresses neutral current events in the detector that produce $\pi^0$'s,
the principle and pernicious background to $\nu_\mu \rightarrow \nu_e$ measurements.
At this energy, the electron-like charged current quasielastic events are straightforwardly
separated from the muon-like events in an ultra-large Cherenkov
detector, as has been well-established by past experiments \cite{MB,
  SK}.      Second, while most low energy conventional beams are produced via targeting off axis,
which yields a narrow band beam which limits reach in $\delta$,  the
low energy of the ESS beam provides low proton energy and produces a
wide-band beam.   Third, if the 540 km ESS baseline is used, then the resulting
energy distributuion of the flux allows forthe  study of the second
oscillation maximum.  The $CP$ violating asymmetry is
significantly larger at the second maximum than at the first maximum,
enhanced by the large value of $\theta_{13}$.  In contrast, the LBNE and T2HK
designs, which were set before $\theta_{13}$ was measured, were chosen
to be most sensitive to the first oscillation maximum.

The alternative approach to measuring 
$\delta$ is to measure the change induced in the oscillation wave as a
function of $L/E$.     The first proposal to pursue this method has been
developed by the DAE$\delta$ALUS collaboration \cite{DAEdALUS}.
This experiment uses cyclotrons at three sites to produce identical
neutrino fluxes from the decay-at-rest of pions and muons.  Events
from the near cyclotron site allows constraint of the initial flux. 
The middle and far sites then allow the shape of the oscillation wave
to be
accurately measured. 
The useful flux from
decay at rest beams range from 20 to 50 MeV, and thus this is a very
short baseline experiment, with the sites located at $<1.5$, 8 and 20 km.  A $\bar
\nu_\mu \rightarrow \bar \nu_e$ signal can be detected in a large
detector with free proton targets (water or scintillation) via  
inverse $\beta$ decay (IBD), $\nu_e + p \rightarrow e^+ n$).  In the
case of a water detector, in order to observe the neutron capture,
Gd-doping, as is presently done in the EGADS experiment \cite{EGADS} 
is required.    Thus, this experiment could use the same detector as
is planned for ESS.     Other detectors under consideration are HyperK
(water) \cite{T2HK}
and LENA (scintillator)\cite{LENA}.

Fig.~\ref{ESS}, red, shows the DAE$\delta$ALUS capability for a 10
year run \cite{DAEdALUS}.     While this sensitivity was calculated for running
DAE$\delta$ALUS with the Hyper-K design,   the result when paired with
ESS will be very similar. This capability is similar to that of the
ESS proposal,
and both approach the measurement of $\delta$ in the quark sector
(gray band).   
The combination can take the measurement of $\delta$ beyond a simple
measurement, to a strong test of the potential presence of non-standard interactions.

\section{Is There Direct Evidence of New Forces or Particles?}

This is the question that unites physicists across many subfields.
In the discussion, we have already highlighted several neutrino 
experiments where new, non-standard forces may be observed.
So here, we consider the potential to directly produce and
observe new particles.    Direct production of new particles 
has been widely regarded
as the preserve of the highest energy scale experiments.   However,
new developments in dark matter \cite{lightdm1, lightdm2} and sterile neutrino 
studies have recently sparked interest at high-intensity, low energy experiments. 

As an example, consider light (0.1 to 10 eV) sterile neutrinos.
Several anomalies have motivated the search for light sterile
neutrinos with mass $\sim$1~eV.  The results arise from short baseline
accelerator, reactor and source experiments
\cite{MB, Lassere,SAGE3, GALLEX3}, and include
both neutrinos and anti-neutrino scattering,  electron and muon
flavors, and more than two orders of magnitude in energy range.
Models which introduce one, two or three sterile neutrinos,
referred to as ``3+1", ``3+2", or ``3+3,'' respectively, have been
introduced to explain the data \cite{Sorel:2003hf}.  Global fits have identified ranges in this extended parameter space where
the anomalies can be reconciled in the 3+2 and 3+3
cases \cite{sterile2013}.

\begin{figure}[tb]
\begin{center}
\includegraphics[scale=.6]{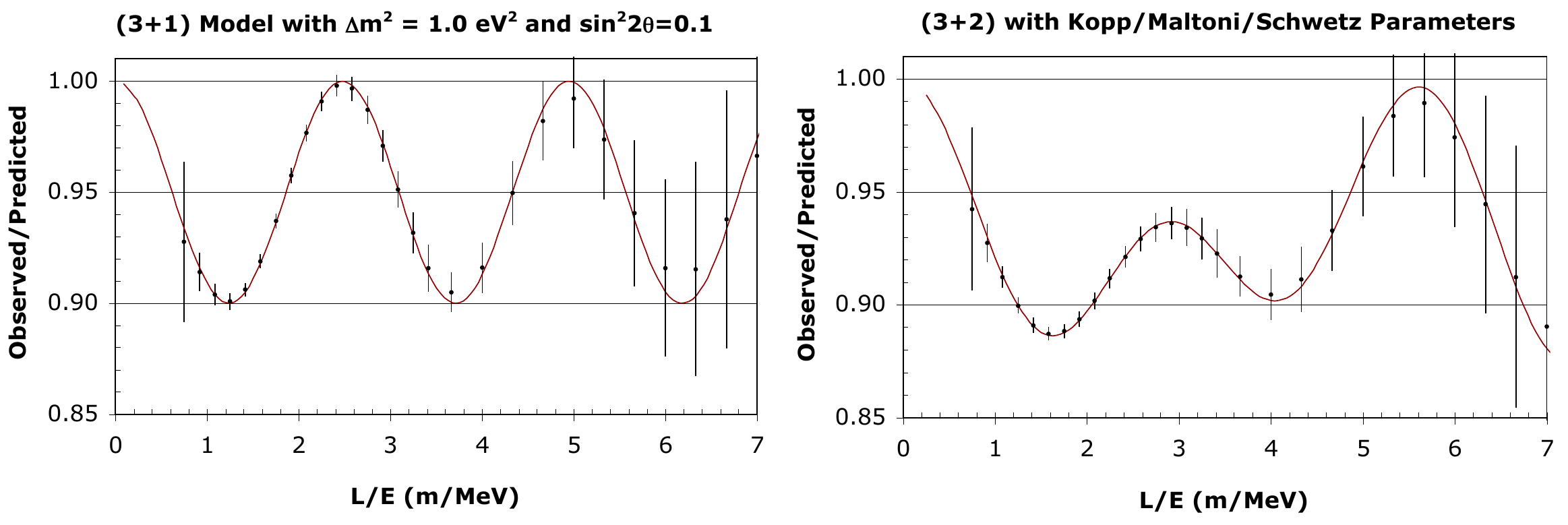}
\end{center}
\vspace{-.5cm}
\caption{Example data sets for 5 years of running for 3+1 (left) and 3+2
  (right) oscillation scenarios for IsoDAR running at KamLAND.
\label{waves}}
\end{figure}

Until now, all of these measurements have been made at specific $L$
values and with rather limited ranges in $E$.
An important goal of the next generation of these searches must be
``oscillimetry,''  where the
oscillation curve is traced in $L/E$ space in a single experiment. This is the only way to
clearly establish that these anomalies arise from oscillations rather
than from some other non-standard, or indeed some unexpected standard, effect.
Sufficient
sensitivity to make a definitive $>5\sigma$ statement is required. 

The experiments that can make decisive measurements are based
on pion/muon or isotope decay-at-rest (DAR) sources. The DAE$\delta$ALUS experiment
described above 
could be used to generate a pion DAR beam for such a
measurement.  However, a proposal which uses the DAE$\delta$ALUS
injector cyclotron,
called IsoDAR, can come on line much more quickly and produce a
definitive result.    This design uses protons from the injector
cyclotron to produce neutrons, which then capture on $^7$Li  to
generate an isotope DAR source. Positioned next to a kiloton-scale scintillator detector such as
KamLAND, one can search for sterile
neutrinos by observing a deficit of antineutrinos as a function of the
distance $L$ and antineutrino energy $E$ across the 
detector\cite{Bungau:2012ea}.

Specifically, the proposed IsoDAR target will be be placed 16 m from
the center of the KamLAND detector.
The antineutrinos propagate a distance of 9.5~m, through a combination of rock, outer muon
veto, and buffer liquid, to the active scintillator volume defined by a
6.5 m radius
nylon balloon.   The antineutrinos are then detected via the IBD
interaction.  The excellent energy and position resolution of KamLAND
leads to a well-reconstructed $L/E$ for the event.  Example data sets
for favored 3+1 and 3+2 sterile 
neutrino parameters are shown in Fig.~\ref{waves}.

\begin{figure}[tb]
\begin{center}
\includegraphics[scale=.6]{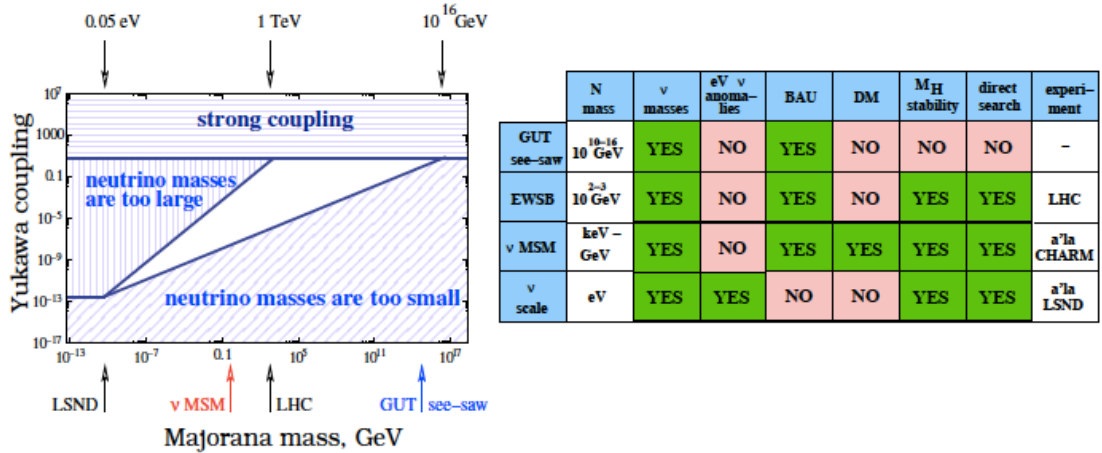}
\end{center}
\vspace{-5.5cm}
\caption{Allowed combinations of sterile neutrino masses and Yukawa
  couplings in see-saw models \cite{rev4}.
\label{sterilesum}}
\end{figure}

The production of sterile neutrinos connects directly back to the LHC
physics.   The previous discussion has considered heavy neutrinos in
the $\sim$ eV range.   However, much heavier neutrinos, in the 100s of
GeV range, can arise from certain see-saw models and loop models
\cite{seesaw}.  
Fig.~\ref{sterilesum}, from Ref. \cite{rev4}, presents possible
allowed masses and 
Yukawa couplings of sterile neutrinos
within seesaw models. The right panel summarizes the role that these
sterile neutrinos may play in solving Standard Model problems and 
Beyond Standard Model anomalies.  The final column shows the preferred
type of experiment to address the sterile neutrino, underlying the
very strong connection between the experiments discussed in this paper and
the LHC.

\section{Conclusion}

This review has highlighted the strong intellectual ties between 
LHC and neutrino experiments.  We have explored the intellectual
overlap through three example questions: What is the origin of mass?
What is the meaning of flavor? Is there direct evidence of new forces
or particles?   However, these are only a few of the questions which
bridge the two communities.    The richness of particle physics as a
field  can be seen by the way these two very different approaches to
experiments  push the community toward new ideas  and questions.


\section*{References}
\begin{thebibliography}{10}

\bibitem{rev1}   C.~Mariani,
  Mod.\ Phys.\ Lett.\ A {\bf 27}, 1230010 (2012)
  [arXiv:1201.6665 [hep-ex]].

\bibitem{rev2}   G.~J.~Feldman, J.~Hartnell and T.~Kobayashi,
  Adv.\ High Energy Phys.\  {\bf 2013}, 475749 (2013)
  [arXiv:1210.1778 [hep-ex]].

\bibitem{rev3}        B.~Schwingenheuer,
  Annalen Phys.\  {\bf 525}, 269 (2013)
  [arXiv:1210.7432 [hep-ex]].


\bibitem{rev4}    K.~N.~Abazajian, {\it et al.},
 ``Light Sterile Neutrinos: A White Paper,''
  arXiv:1204.5379 [hep-ph].

\bibitem{rev5}   A.~Karle,
  Nucl.\ Phys.\ Proc.\ Suppl.\  {\bf 235-236}, 364 (2013)
  [arXiv:1210.2058 [astro-ph.IM]].

\bibitem{LEP}   S.~Schael {\it et al.}  [ALEPH and DELPHI and L3 and OPAL and SLD and LEP Electroweak Working Group and SLD Electroweak Group and SLD Heavy Flavour Group Collaborations],
  Phys.\ Rept.\  {\bf 427}, 257 (2006)
  [hep-ex/0509008].


\bibitem{mixresults}  M.~C.~Gonzalez-Garcia, M.~Maltoni, J.~Salvado and T.~Schwetz,
  JHEP {\bf 1212}, 123 (2012)
  [arXiv:1209.3023 [hep-ph]].


\bibitem{beta}    C.~Weinheimer and K.~Zuber,
  Annalen der Physik, {\bf 525}, 565 (2013)
  [arXiv:1307.3518 [hep-ex]].

\bibitem{atmospheric}
  T.~Kajita,
  Adv.\ High Energy Phys.\  {\bf 2012}, 504715 (2012).
\bibitem{solar}    M.~Nakahata,
  Lect.\ Notes Phys.\  {\bf 367}, 49 (1990).


\bibitem{seesaw}    R.~N.~Mohapatra,
  J.\ Phys.\ Conf.\ Ser.\  {\bf 408}, 012005 (2013).


\bibitem{KATRIN}    M.~Sturm [KATRIN Collaboration],
  PoS DSU {\bf 2012}, 037 (2012).

\bibitem{P8}   J.~A.~Formaggio [Project 8 Collaboration],
  Nucl.\ Phys.\ B, Proc.\ Suppl.\ 229-232 {\bf 2012}, 371 (2012)
  [arXiv:1101.6077 [nucl-ex]].


\bibitem{Leptogen}     S.~Pascoli, S.~T.~Petcov and A.~Riotto,
  Phys.\ Rev.\ D {\bf 75}, 083511 (2007)
  [hep-ph/0609125].

\bibitem{Yeh}   M.~Yeh, A.~Garnov and R.~L.~Hahn,
  Nucl.\ Instrum.\ Meth.\ A {\bf 578}, 329 (2007).


\bibitem{Klein}  J. Klein, private communication.

\bibitem{pinguwhitepaper}    T.~IceCube [PINGU Collaboration],
  ``PINGU Sensitivity to the Neutrino Mass Hierarchy,''
  arXiv:1306.5846 [astro-ph.IM].


\bibitem{paper16}     P.~A.~R.~Ade {\it et al.}  [Planck Collaboration],
  ``Planck 2013 results. XVI. Cosmological parameters,''
  arXiv:1303.5076 [astro-ph.CO].


\bibitem{ckm95}     L. Montanet, {\it et al.} [Particle Data Group],
The Review of Particle Physics, Phys. Rev. D50 (1995) 1173.


\bibitem{DC}    Y.~Abe {\it et al.}  [DOUBLE-CHOOZ Collaboration],
  Phys.\ Rev.\ Lett.\  {\bf 108}, 131801 (2012)
  [arXiv:1112.6353 [hep-ex]].


\bibitem{DB}    F.~P.~An {\it et al.}  [Daya Bay Collaboration],
  Chin.\  Phys.\ C {\bf 37}, 011001 (2013)
  [arXiv:1210.6327 [hep-ex]].

\bibitem{RENO}   J.~K.~Ahn {\it et al.}  [RENO Collaboration],
  Phys.\ Rev.\ Lett.\  {\bf 108}, 191802 (2012)
  [arXiv:1204.0626 [hep-ex]].



\bibitem{Thiago}    T.~J.~C.~Bezerra, H.~Furuta, F.~Suekane and T.~Matsubara,
  Phys.\ Lett.\ B {\bf 725}, 271 (2013)
  [arXiv:1304.6259 [hep-ex]].

\bibitem{MINOS}    P.~Adamson {\it et al.}  [MINOS Collaboration],
  Phys.\ Rev.\ Lett.\  {\bf 110}, 251801 (2013)
  [arXiv:1304.6335 [hep-ex]].



\bibitem{Huber}  P. Huber,  Talk presented on Friday, ``CP violation reach in the next
decade and beyond,'' presented at Snowmass at the Mississippi
July 26 – August 6, 2013, Minneapolis.

\bibitem{mine}     T.~Enqvist, A.~Mattila, V.~Fohr, T.~Jamsen, M.~Lehtola, J.~Narkilahti, J.~Joutsenvaara and S.~Nurmenniemi {\it et al.},
  Nucl.\ Instrum.\ Meth.\ A {\bf 554}, 286 (2005)
  [hep-ex/0506032].


\bibitem{MEMPHYS}
  L.~Agostino {\it et al.}  [MEMPHYS Collaboration],
  JCAP {\bf 1301}, 024 (2013)
  [arXiv:1206.6665 [hep-ex]].

\bibitem{ESS}
  E.~Baussan, M.~Dracos, T.~Ekelof, E.~F.~Martinez, H.~Ohman and N.~Vassilopoulos,
  arXiv:1212.5048 [hep-ex].


\bibitem{LBNE} 
  C.~Adams {\it et al.}  [LBNE Collaboration],
  arXiv:1307.7335 [hep-ex].


\bibitem{T2HK}     E.~Kearns {\it et al.}  [Hyper-Kamiokande Working Group Collaboration],
  ``Hyper-Kamiokande Physics Opportunities,''
  arXiv:1309.0184 [hep-ex].


\bibitem{MB}    A.~A.~Aguilar-Arevalo {\it et al.}  [MiniBooNE Collaboration],
  Nucl.\ Instrum.\ Meth.\ A {\bf 599}, 28 (2009)
  [arXiv:0806.4201 [hep-ex]].


\bibitem{SK}     M.~Fechner {\it et al.}  [Super-Kamiokande Collaboration],
  Phys.\ Rev.\ D {\bf 79}, 112010 (2009)
  [arXiv:0901.1645 [hep-ex]].

\bibitem{EGADS}    M.~R.~Vagins,
  Nucl.\ Phys.\ Proc.\ Suppl.\  {\bf 229-232}, 325 (2012).


\bibitem{LENA}   M.~Wurm {\it et al.}  [LENA Collaboration],
  Astropart.\ Phys.\  {\bf 35}, 685 (2012)
  [arXiv:1104.5620 [astro-ph.IM]].

\bibitem{DAEdALUS}    C.~Aberle, A.~Adelmann, J.~Alonso, W.~A.~Barletta, R.~Barlow, L.~Bartoszek, A.~Bungau and A.~Calanna {\it et al.},
  arXiv:1307.2949 [physics.acc-ph].

\bibitem{lightdm1}  C.~Boehm, P.~S.~B.~Dev, A.~Mazumdar and E.~Pukartas,
  JHEP {\bf 1306}, 113 (2013)
  [arXiv:1303.5386 [hep-ph]].


\bibitem{lightdm2}   Y.~Zhang, X.~Ji and R.~N.~Mohapatra,
  ``A Naturally Light Sterile neutrino in an Asymmetric Dark Matter Model,''
  arXiv:1307.6178 [hep-ph].

\bibitem{MB}     J.~M.~Conrad, W.~C.~Louis and M.~H.~Shaevitz,
  Ann.\ Rev.\ Nucl.\ Part.\ Sci.\  {\bf 63}, 45 (2013)
  [arXiv:1306.6494 [hep-ex]].


\bibitem{Lassere}  G.~Mention, {\it et al.}, 
  Phys.\ Rev.\  D {\bf 83}, 073006 (2011)
  [arXiv:1101.2755 [hep-ex]].


\bibitem{SAGE3}   J.~N.~Abdurashitov {\it et al.}  [SAGE Collaboration],
  Phys.\ Rev.\  C {\bf 80}, 015807 (2009)
  [arXiv:0901.2200 [nucl-ex]].


\bibitem{GALLEX3}   F.~Kaether, W.~Hampel, G.~Heusser, J.~Kiko and T.~Kirsten,
  Phys.\ Lett.\  B {\bf 685}, 47 (2010)
  [arXiv:1001.2731 [hep-ex]].


\bibitem{Sorel:2003hf}   
   M.~Sorel, J.~M.~Conrad and M.~Shaevitz,
  Phys.\ Rev.\  D {\bf 70}, 073004 (2004)
  [arXiv:hep-ph/0305255].
 

\bibitem{sterile2013}    J.~M.~Conrad, C.~M.~Ignarra, G.~Karagiorgi, M.~H.~Shaevitz and J.~Spitz,
  Adv.\ High Energy Phys.\  {\bf 2013}, 163897 (2013)
  [arXiv:1207.4765 [hep-ex]].


\bibitem{Bungau:2012ea}
  A.~Bungau, A.~Adelmann, J.~R.~Alonso, W.~Barletta, R.~Barlow, L.~Bartoszek, L.~Calabretta and A.~Calanna {\it et al.},
  Phys.\ Rev.\ Lett.\  {\bf 109}, 141802 (2012)
  [arXiv:1205.4419 [hep-ex]].


\end{thebibliography}
\end{document}